\documentclass[aps,prd]{revtex4}
\newcommand{\et}{\tilde{e}}
\newcommand{\At}{\tilde{A}}
\newcommand{\rt}{\tilde{\rho}}
\begin{document}
\title{A Remark on Vector Meson Dominance, Universality and the Pion
  Form Factor} 

\author{ Alfonso R. Zerwekh}
\email[]{alfonsozerwekh@uach.cl}
 \affiliation{Instituto de F{\'{\i}}sica, Facultad de Ciencias \\
Universidad Austral de Chile \\ Casilla 567, Valdivia, Chile}

\begin{abstract}
 In this paper, we address the universality problem in the mass-mixing
 representation of vector meson dominance. First we stress the
 importance of using physical (mass eigenstate) fields in order to get
 the correct $q^2$ dependence of the pion form factor. Then we
 show that, when a direct coupling of the (proto-)photon to the pions
 is included, it is not necessary to invoke universality. Our method
 is similar to the delocalization idea in some deconstrunction
 theories.         
\end{abstract}

\maketitle

In 1960, Sakurai \cite{Sakurai:1960ju} proposed a theory of strong
interactions based on the idea of local gauge invariance where the
interaction was supposed to be mediated by vector mesons. In this 
scenario, the electromagnetic interaction of hadrons was introduced
through a mixing of the photon and the vector mesons. This idea is
known as Vector Meson Dominance (VMD) \cite{O'Connell:1995wf}. 

Historically, two lagrangian realizations of
VMD have been in use. The first one, due to Kroll, Lee and Zumino
\cite{Kroll:1967it}, describes 
the mixing of the photon and the rho meson through a term of the form:

\begin{equation}
  \label{eq:vmd-1}
  {\cal L}_{\gamma \rho}=\frac{e}{2g_{\rho}}\rho_{\mu \nu}F^{\mu \nu}.
\end{equation}
\noindent
This representations is usually called VMD-1. In general, it is viewed
as the more elegant VMD realization because it is explicitly
consistent with electromagnetic gauge invariance and the pion form
factor calculated from it can be written as:

\begin{equation}
  \label{eq:pion-ff-vmd1}
  F_{\pi}(q^2)=\left[1-\frac{q^2}{q^2-m^2_{\rho}}\frac{g_{\rho \pi
  \pi}}{g_{\rho}}\right] 
\end{equation}
\noindent
which satisfy the condition $F_{\pi}(0)=1$ without any assumption
about the coupling constants $g_{\rho\pi\pi}$ and $g_{\rho}$. The
price to pay is to work with non-diagonal propagators.

The second and, in some sense, more popular  realization (usually
called VMD-2) is based on a mass-mixing term of the form:

\begin{equation}
  \label{eq:vmd-2}
   {\cal L}_{\gamma \rho}=-\frac{e m^2_{\rho}}{g_{\rho}}\rho_{\mu}A^{\mu}.
\end{equation}
\noindent
 This version of VMD is seen as unsatisfactory because the mixing term
 (\ref{eq:vmd-2}) introduce corrections to the photon propagator
 acquiring a non-zero mass. In order to correct this important flaw,
 it is necessary to add to the lagrangian a mass term for the
 photon. On the other hand, when the pion form factor is calculated in
 this representation the result obtained is:

 \begin{equation}
   \label{eq:pion-ff-vmd2}
 F_{\pi}(q^2)=-\frac{m^2_{\rho}}{q^2-m^2_{\rho}}\frac{g_{\rho \pi
  \pi}}{g_{\rho}}. 
 \end{equation}

In this case, the condition $F_{\pi}(0)=1$ is satisfied only if $g_{\rho \pi
  \pi}=g_{\rho}$. This is the universality condition. It is fair to
  say that both version of VMD are seen as equivalent in the limit of
  universality. 

In the rest of this paper we discuss some ideas in order to evade the
need of universality. We consider a simplified model in which the
(neutral) rho meson is treated as an abelian field and we do not take
into account the charged rho mesons in order to concentrate our
attention on the main features of the mechanism. A brief discussion
about how to include the charged rho mesons is presented in the
Appendix.

We start by writing down the VMD-2 lagrangian for the vector sector:

\begin{eqnarray}
  \label{eq:lagrangian}
  {\cal L}&=&-\frac{1}{4}\tilde{F}_{\mu \nu}\tilde{F}^{\mu \nu}
           -\frac{1}{4}\rt_{\mu \nu}\rt^{\mu \nu}
           +\frac{1}{2}m^2_{\rho}\rt_{\mu}\rt^{\mu}
           -\frac{\et m^2_{\rho}}{g_{\rho}}\rt_{\mu}\At^{\mu} 
\nonumber \\
        & &+\frac{1}{2}\left(\frac{\et
  m_{\rho}}{g_{\rho}}\right)^2\At_{\mu}\At^{\mu}. 
\end{eqnarray}
We will call the fields $\rt_{\mu}$ and $\At_{\mu}$  the proto-rho and
the proto-photon, respectively. This lagrangian is gauge invariant if
the proto-photon and the proto-rho transform under $U(1)_{\mbox{\tiny{EM}}}$ as:

\begin{eqnarray}
  \label{eq:trans}
\delta \At_{\mu}&=&\frac{1}{\et}\partial_{\mu}\Lambda \\
\delta \rt_{\mu}&=&\frac{1}{g_{\rho}}\partial_{\mu}\Lambda  
\end{eqnarray}

 At this point, it is necessary to remark
that neither the proto-rho nor the proto-photon are mass eigenstate,
and hence, they are not physical fields. The importance of using a
physical basis has been recognized and advocated by other authors
\cite{Schechter:1986vs}. In fact, in a bit different context, the use
of the physical 
basis has help to enlighten the interaction between a color octet
technirho and gluons \cite{Zerwekh:2001uq}. In our case, electromagnetic
gauge invariance enforces the mass matrix to have a null determinant, and
hence, it implies that the physical photon is massless. When the mass matrix is
diagonalized we find that the physical rho and photons fields are:

\begin{eqnarray}
  \label{eq:physcal_fields}
  A_{\mu}&=& \At_{\mu}\cos\alpha + \rt_{\mu}\sin\alpha\\
 \rho_{\mu}&=&-\At_{\mu}\sin\alpha + \rt_{\mu}\cos\alpha
\end{eqnarray}
where
$$
  \cos\alpha=\frac{g_{\rho}}{\sqrt{\et^2+g^2_{\rho}}}
$$
and   
$$
  \sin\alpha=\frac{\et}{\sqrt{\et^2+g^2_{\rho}}}.
$$

Let now turn our attention to the charged pions. They are
described by the lagrangian:

\begin{equation}
  \label{eq:pions}
  {\cal L_{\pi}}=D_{\mu}\pi^{+}D^{\dagger \mu}\pi^{-}-m_{\pi}^2\pi^{+}\pi^{-},
\end{equation}
where
\begin{equation}
  \label{eq:Derivada_Covariante}
  D_{\mu}=\partial_{\mu}+ix\et\At_{\mu}+i(1-x)g_{\rho}\rt_{\mu}
\end{equation}
is the more general covariant derivative we can form with the fields
$\At_{\mu}$ and $\rt_{\mu}$.

Notice that here we slightly deviate from traditional VMD-2 because we
include a direct coupling between the pions and the proto-photon. In
this sense, we are advocating for a partial vector meson
dominance. It is this direct coupling with the proto-photon what will
allow us to abandon universality. Some indications of a deviation from
complete vector meson
dominance were already communicated and discussed in \cite{Schechter:1986vs}.
Nevertheless, as far as we know this relation between universality and
a direct coupling of the proto-photon to pions, have not been
discussed before. On
the other hand, the $x$ variable plays a r\^ole similar to
delocalization parameters in some deconstruction models
\cite{Chivukula:2005bn}. 

In terms of the physical fields the covariant derivative can be
written as:

\begin{equation}
  \label{eq:Derivada_Covariante_Fisica}
  D_{\mu}=\partial_{\mu}+ieA_{\mu}+ig_{\rho\pi\pi}\rho_{\mu}
\end{equation}
where $e=\et\cos\alpha=g_{\rho}\sin\alpha$ is the electric charge of the
positron, and $g_{\rho\pi\pi}$ can be expressed as:

\begin{equation}
  \label{eq:grpp}
  g_{\rho\pi\pi}=g_{\rho}\cos\alpha\left(1-\frac{x}{\cos^2\alpha}\right).
\end{equation}

In this context universality means
$g_{\rho\pi\pi}=g_{\rho}\cos\alpha$, what only happens  when
$x=0$. 

On the other hand, we must avoid a direct coupling of the proto-rho with
leptons (leptons do not interact strongly) and hence they will be
described by the lagrangian:

\begin{eqnarray}
  \label{eq:leptons}
  {\cal
  L}&=&\bar{\psi}\gamma^{\mu}\left(i\partial_{\mu}+\et\At_{\mu}\right)\psi
  \nonumber \\
   &=&\bar{\psi}\gamma^{\mu}\left(i\partial_{\mu}+eA_{\mu}-e\tan\alpha\rho_{\mu}\right)\psi
\end{eqnarray}

With these ingredients, we obtain the following expression for the
amplitude  of the process $e^+ e^- \rightarrow \pi^+ \pi^-$:

\begin{equation}
  \label{eq:amplitude}
{\cal
  M}=-ie^2\bar{v}\gamma_{\mu}u(k_1-k_2)^{\mu}\left[1-\frac{q^2}{q^2-M_{\rho}}
 \left(1-\frac{x}{\cos^2\alpha} \right) \right]   
 \end{equation}

Hence, the pion form factor may be written as:

\begin{equation}
  \label{eq:F(q2)}
  F(q^2)=\left[1-\frac{q^2}{q^2-M_{\rho}}
 \left(1-\frac{x}{\cos^2\alpha} \right) \right]
\end{equation}
or, in  a more familiar way:
\begin{equation}
  \label{eq:F(q2)v2}
  F(q^2)=\left[1-\frac{q^2}{q^2-M_{\rho}}
 \frac{g_{\rho\pi\pi}}{g_{\rho}\cos\alpha} \right]
\end{equation}

The pion form factor we obtained is similar to the one obtained using
VMD-1: it has a correct behavior for $q^2=0$ and it does not depend on
universality. The first feature is a consequence of having used
physical fields while the second has its roots in the direct coupling
of the proto-photon with pions.

The values of the mixing angle $\alpha$ and the $x$ parameter can be
obtained from experiment. The partial decay widths $\Gamma(\rho
\rightarrow e^+ e^-)$ and $\Gamma(\rho \rightarrow \pi^+ \pi^-)$ can
be written as:

\begin{eqnarray}
 \Gamma(\rho \rightarrow e^+ e^-)&=& \frac{1}{3} \alpha_{\mbox{{\tiny
 EM}}} \tan^2\alpha M_{\rho}\\
 \label{eq:gammas}
 \Gamma(\rho \rightarrow \pi^+ \pi^-)&=&\frac{\alpha_{\mbox{{\tiny
 EM}}}}{12}\frac{g_{\rho\pi\pi}^2}{g_{\rho}^2}\frac{
 M_{\rho}}{\sin^2\alpha}\left( 1-\frac{4m_{\pi}^2}{
 M_{\rho}^2}\right)^{3/2} 
\end{eqnarray}
where $\alpha_{\mbox{{\tiny EM}}}$ is the electromagnetic
fine-structure constant. Using the experimental values $\Gamma(\rho
\rightarrow e^+ e^-)=6.85$ keV and $\Gamma(\rho \rightarrow \pi^+
\pi^-)=146.4$ MeV \cite{pdg} we obtain $\tan\alpha=0.0603$, $x = -0.177$,
$g_{\rho\pi\pi}=5.92$ and $g_{\rho}=5.03$.

In conclusion, we have constructed a VMD-2-like lagrangian which
correctly describes the pion form factor without needing the
universality hypotheses. The main ingredients of our approach are a
strong use of electromagnetic gauge invariance, 
the use of physical fields and to allow a direct coupling of the
proto-photon with the pions implementing a partial domination of the
rho vector meson.

\appendix

\section*{Appendix}

When we wrote down the lagrangian of the vector sector of VMD-2, we
choosed a representation where both, the proto-rho and the
proto-photon, transform like gauge fields under
$U(1)_{\mbox{\tiny{EM}}}$. Nevertheless, in the physical basis, only
the photon transform as a gauge field while the physical rho
transforms trivially because it is neutral. On the other hand, the
charged rho transform as:

\begin{equation}
  \label{eq:charged_rho_transformation}
  \rho^{\pm} \rightarrow e^{\pm i \Lambda} \rho^{\pm}.
\end{equation}

Of course, the lagrangian

\begin{equation}
  \label{eq:charged_rho_lagrangian}
{\cal L}=
-\frac{1}{4}(D_{\mu}\rho^{+}_{\nu}-D_{\nu}\rho^{+}_{\mu})(D^{\mu}\rho^{-\nu}-D^{\nu}\rho^{-\mu}) 
+ \tilde{M_{\rho}}\rho^{+}_{\mu}\rho^{-\mu},   
\end{equation}
where $D_{\mu}$ is the usual covariant derivative, is gauge
invariant. To this lagrangian can be added all the interaction terms
consistent with gauge invariance and the global symmetries we want to
implement such as isospin invariance.


\end{document}